\documentstyle{article}
\input boxedeps.tex
\SetRokickiEPSFSpecial  
\HideDisplacementBoxes
\begin{document}

\mbox{}\hfill MPIH--V18--1996\\
\mbox{}\hfill hep-ph/9605409\\

\begin{center}
{\LARGE  Convergence of Discretized Light Cone Quantization in 
the small mass limit\\}
\vspace{10mm}

{\bf Brett van de Sande}\\
\vspace{10mm}

{\em
Max Planck Institut f\"ur Kernphysik,\\
Postfach 10.39.80, D-69029 Heidelberg, Germany }
\vspace{10mm}

\end{center}

\begin{abstract}
I discuss the slow convergence of Discretized Light Cone Quantization
(DLCQ) in the small mass limit and suggest a solution.
\end{abstract}

\baselineskip .25in
\section{Introduction}

``Light-front field theory'' is an attempt to apply 
Hamiltonian techniques to relativistic field theory;
the idea is to quantize a theory on a surface of constant
$x^+ = (x^0+x^3)/\sqrt 2$ and calculate
eigenstates of the invariant mass squared operator $P^\mu P_\mu$.
In order to partially regulate a theory, one can simply impose 
(anti-)periodic boundary conditions in the $x^-$ coordinate;
this is known as ``Discretized Light Cone Quantization'' (DLCQ)
\cite{dlcq} since the momenta conjugate to $x^-$ become discrete.  
One can use DLCQ to construct a basis of states for use in
numerical calculations.

Why would one want to use DLCQ in a large scale numerical 
calculation?
Ideally, one would employ a basis of polynomial-type wavefunctions
to solve the bound state equation \cite{harada}.  
With such a basis, convergence
of the eigenvalues is typically exponentially fast.
However, the practitioner faces several difficulties in applying
this approach to a many particle calculation.
In general, the resulting basis is not orthogonal and
the resulting Hamiltonian matrix is no longer very sparse.
Second, matrix elements become difficult to compute
since they involve complicated integrals.
It is not clear that the faster convergence of such a basis can 
compensate for these added difficulties.

In contrast, a DLCQ basis is orthogonal and
the corresponding Hamiltonian matrix is sparse.
Also, matrix elements are easily calculated; no integrals 
need to be evaluated.  This is important since the time required to 
calculate matrix elements is currently 
the limiting factor in large calculations.
Finally, as a regulator, DLCQ has an appealing physical interpretation
in terms of (anti-)periodic boundary conditions.

However, DLCQ has one distinct disadvantage:
it converges quite slowly in the small mass limit.
This is especially of concern for the transverse lattice
\cite{bardeen,dalley} where the continuum limit is precisely 
the small mass limit. In particular, the lowest eigenvalue 
and measurements of the string tension are strongly affected by this 
slow convergence.
As we shall see, the source of this slow convergence 
comes from the end-point behavior of the wavefunction which
is handled poorly by DLCQ.  Using the 't Hooft equation as an
example, we will study this behavior and suggest a
solution that can be easily applied to other theories
and large numerical calculations.

\section{The 't Hooft equation}

\begin{figure}
\centering
\BoxedEPSF{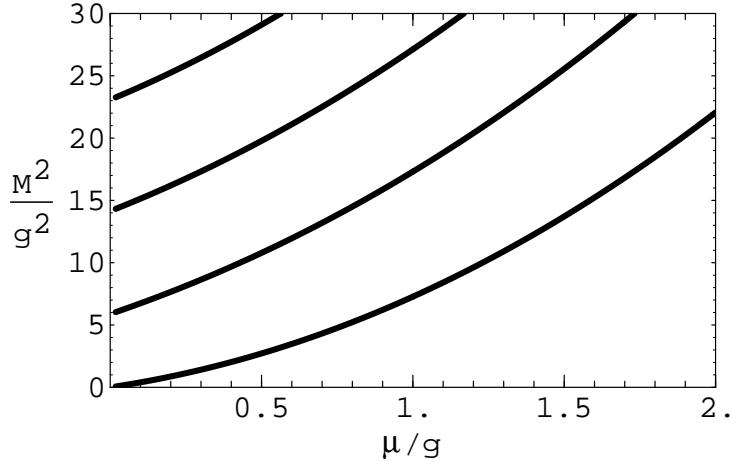 scaled 1200}
\caption{
Lowest eigenvalues of the 't Hooft equation as
a function of mass $\mu$ calculated using
a polynomial wavefunction basis~\protect\cite{harada}. \label{spec}}
\end{figure}

The 't Hooft equation \cite{thooft} 
is the bound state equation for two quarks in the large $N$ limit
of (1+1)-dimensional QCD:
\begin{equation}
   M^2 \Psi(x) = \mu^2 \Psi(x)\left(\frac{1}{x}+\frac{1}{1-x}\right)
    + g^2 \int_0^1 dy \, \frac{\Psi(x)-\Psi(y)}{\left(x-y\right)^2}
   \label{thooft}
\end{equation}
where $M$ is the Lorentz invariant mass eigenvalue associated
with eigenfunction $\Psi(x)$.
A Cauchy principle value prescription is assumed throughout.
The spectrum is shown in Fig.~\ref{spec}.
Following 't Hooft, let us analyze the endpoint behavior of the wavefunction.
Assume 
\begin{equation}
    \lim_{x \to 0} \Psi(x) \propto x^\beta \; .
\end{equation}
For small $x$,  the bound state equation (\ref{thooft}) becomes
\begin{equation}
   0 = x^{\beta-1}\mu^2-g^2 x^{\beta-1}-
     g^2 \int_0^1 dy\, \frac{y^\beta}{(x-y)^2}\left(1+O(y)\right) + 
     O\!\left(x^\beta\right) \; .
\end{equation}
Substituting $z=y/x$, we obtain, in the $x \to 0$ limit,
\begin{eqnarray}
   0 &=& \frac{\mu^2}{g^2}-1- \int_0^{1/x} dz\, 
         \frac{z^\beta}{(1-z)^2}\label{int} \\ 
     &=& \frac{\mu^2}{g^2}-1+ \pi \beta \cot(\pi \beta) \label{beta} \; .
\end{eqnarray}
Convergence of the integral (\ref{int}) implies $0<\beta<1$.
For small mass $\mu$ 
\begin{equation}
  \beta = \frac{\mu \sqrt{3}}{g \pi}\,\left(1-\frac{\mu^2}{10 g^2}
        + \cdots\right) \; .
\end{equation}
In addition, one can derive an expression for the lowest 
eigenvalue in this limit
\begin{equation}
     M^2 = \frac{2 \pi g \mu}{\sqrt 3} + O\!\left(\mu^2\right) \; .
\end{equation}

\section{Analysis and discussion}

Exact solutions of the 't Hooft equation (\ref{thooft}) are not known and
one must resort to numerical techniques.  For the purpose of 
comparison, we will introduce a basis of wavefunctions of the form 
$x^\beta (1-x)^\beta P_n(x)$ where $P_n(x)$ is some polynomial
of order $n$ and $\beta$ is given by Eqn.~(\ref{beta}). 
Such a basis provides very quick numerical convergence
\cite{harada}.  Later, when I quote ``exact eigenvalues,'' I will
use results from such a calculation.

\begin{figure}
\centering
\BoxedEPSF{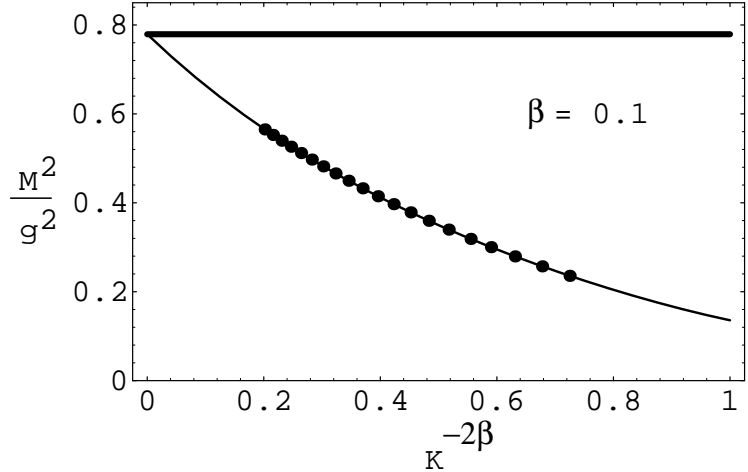 scaled 1200}
\caption{
Convergence of the lowest eigenvalue with $K$ for
$\beta=0.1$ ($\mu=0.181981$).  
The dots are from DLCQ and the
heavy line is the exact result. \label{fig1}}
\end{figure}

Alternatively, we can use DLCQ to solve the bound state equation.
We discretize the momentum fraction $x$ as $i/K$ where
$i\in \{1/2,3/2,\ldots,K-1/2\}$ for integer-valued ``harmonic
resolution'' $K$.  Physically, this discretization corresponds 
to applying anti-periodic boundary conditions in the $x^-$ 
direction where $K\to \infty$ is the continuum limit.  
The DLCQ version of the bound state equation (\ref{thooft}) is 
\begin{equation}
  M^2 \Psi_i = \mu^2 K \Psi_i \left(\frac{1}{i}+\frac{1}{K-i}\right)
    + g^2 K \sum_{j \neq i} \frac{\Psi_i-\Psi_j}{(i-j)^2}  \; .
    \label{dlcq}
\end{equation}
Although a proof is lacking, numerical evidence suggests 
that the DLCQ eigenvalues converge as $1/K^{2 \beta}$.  
For example, in Fig.~\ref{fig1}, we plot the lowest DLCQ eigenvalue 
as a function of $K$ together with a numerical fit to 
\begin{equation}
M^2 = 0.779141 - \frac{1.25943}{ K^{2 \beta}} + \frac{1.15609}{ K^{4 \beta}}
      - \frac{1.01505}{ K^{6 \beta}} +\cdots
\end{equation}
where the ``exact eigenvalue'' (from a polynomial basis calculation) 
is $M^2=0.77914$. 

Note that the DLCQ bound state equation does not make use of
our knowledge of the endpoint behavior of the wavefunction.
This is, in fact, the source of the slow DLCQ convergence.
We can include the endpoint behavior by 
adding and subtracting a term from the 't Hooft 
equation:
\begin{equation}
   M^2 \Psi(x) = \mu^2 \Psi(x)\left(\frac{1}{x}+\frac{1}{1-x}\right)
    +g^2 \Psi(x) \, {\rm I}(x)
    + g^2 \int_0^1 dy \, \frac{\Psi(x)
    \left(\frac{y(1-y)}{x(1-x)}\right)^\beta-\Psi(y)}{(x-y)^2}
   \label{thooft2}
\end{equation}
where
\begin{equation}
   {\rm I}(x)= \int_0^1 dy\, \frac{1-\left(\frac{y(1-y)}{x(1-x)}\right)^\beta
             }{(x-y)^2} \; .
\end{equation}
The DLCQ version of this equation is
\begin{equation}
  M^2 \Psi_i = \mu^2 K \Psi_i \left(\frac{1}{i}+\frac{1}{K-i}\right)
     + g^2 \Psi_i \, {\rm I}(i/K)
      + g^2 K \sum_{j \neq i} \frac{\Psi_i\left( \frac{j(K-j)}{i(K-i)}
      \right)^\beta-\Psi_j}{(i-j)^2} 
      \label{improved}
\end{equation}
where we evaluate the integral ${\rm I}(x)$ 
exactly.\footnote{In practice, I solve it numerically.}
Based on the endpoint analysis, the integrand
in Eqn.~(\ref{thooft2}) should vanish when both $x$ and $y$ are near 
one of the endpoints.  Thus, the corresponding discretization
(\ref{improved}) should
have small errors in this region.
We will refer to this form of the bound
state equation as ``improved DLCQ.''        

\begin{figure}
\centering
\BoxedEPSF{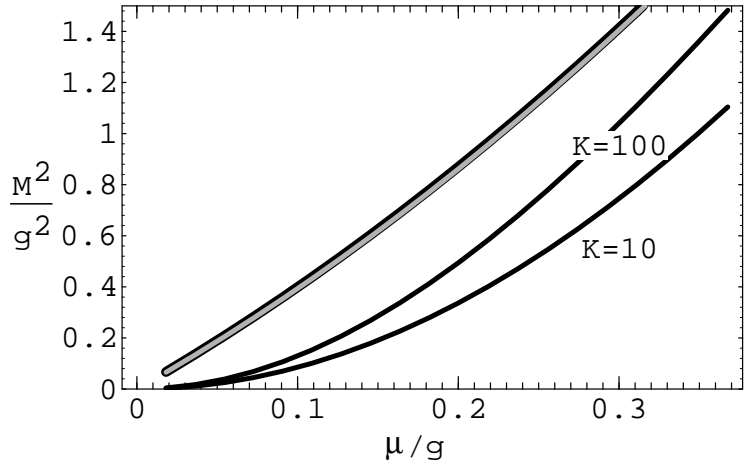 scaled 1200}
\caption{
The lowest eigenvalue as a function of $\mu$.  
The black lines are from DLCQ ($K=10$ and 100) 
and the heavy line is the exact result.  The improved DLCQ
eigenvalue (gray line, $K=10$) is almost identical
to the the exact result. \label{fig3}}
\end{figure}

Let us see how this affects convergence of the spectrum.
In Fig.~\ref{fig3}, we see that the DLCQ results converge
slowly and, in fact, have an incorrect functional form for
$M^2$ versus $\mu$ in the small $\mu$ limit. On the other hand, the improved
DLCQ eigenvalues are nearly identical to the exact result.
However, this absence of finite $K$ corrections is the result 
of a fortuitous accident:
for small $\mu$, the lowest eigenfunction is 
$\Psi(x) \approx x^\beta (1-x)^\beta$.
Thus, the last term of Eqn.~(\ref{improved})
is almost zero for all $i$ and $j$ and {\em all}\/ 
finite $K$ corrections vanish.

Examining an excited state, where the wavefunction has a 
more complicated form, will give us a better picture of
what happens in the generic case.  Ordinarily, DLCQ
eigenvalues converge as $1/K^{2 \beta}$ due to the endpoint
regions $x \approx 0$ and $x\approx 1$ plus $1/K$ errors due to the 
$1/(x-y)^2$ singularity~\cite{hornbostel}.
\begin{figure}
\centering
\BoxedEPSF{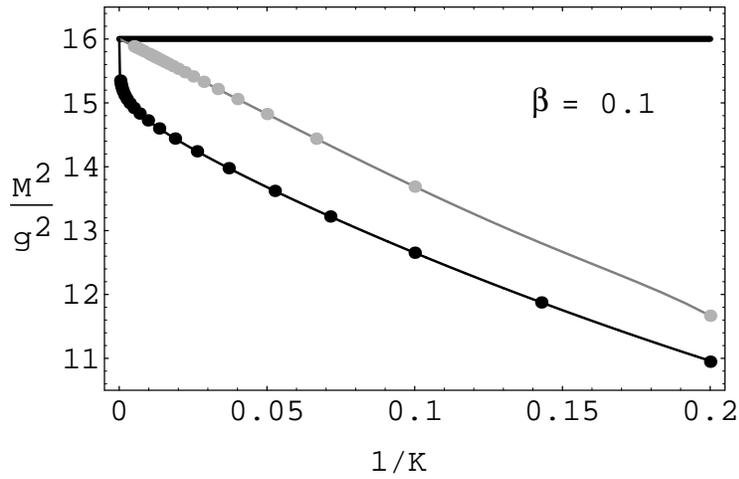 scaled 1200}
\caption{
Convergence of the third lowest eigenvalue with $K$  for
$\beta=0.1$ ($\mu=0.181981$).  
The black dots are for DLCQ; the gray dots are for
improved DLCQ; and the
heavy line is the exact result. \label{fig2}}
\end{figure}
Thus, in Fig.~\ref{fig2}, the DLCQ eigenvalues of the second
excited state are fitted to powers
of $1/K^{2 \beta}$ plus powers of $1/K$,
\begin{equation}
  M^2 = 16.0016 - \frac{3.86429}{K^{2\beta}}+\frac{3.88433}{K^{4\beta}} 
       +\cdots  - \frac{20.7358}{K} + \frac{19.0972}{K^2} \; ,
\end{equation}
while the improved DLCQ eigenvalues are fitted to powers of $1/K$,
\begin{equation}
  M^2 = 16.0193 - \frac{24.7124}{K}  + \frac{14.7355}{K^2} \; .
\end{equation}
The exact result is $M^2=16.0016$.
It is important to note that improved DLCQ removes the 
errors associated with the endpoint regions but does not
significantly affect the $1/K$ errors.
Finally, in Fig.~\ref{fig4} we see how convergence behaves
as a function of $\mu$ for the second excited state. 
\begin{figure}
\centering
\BoxedEPSF{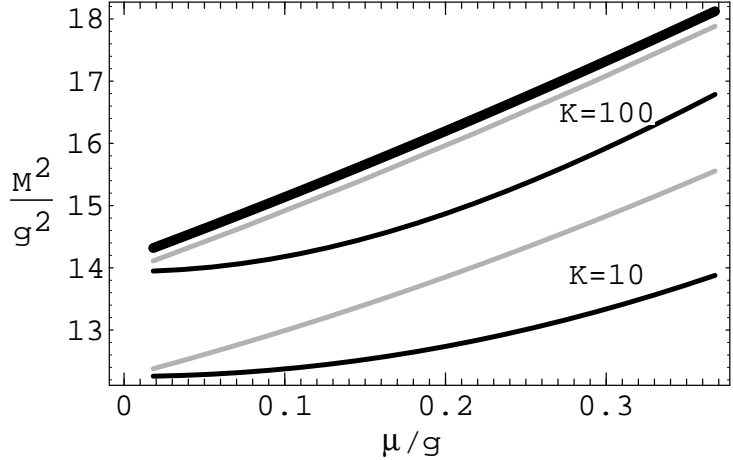 scaled 1200}
\caption{
The third lowest eigenvalue as a function of $\mu$.
The black lines are from DLCQ ($K=10$ and 100);
the gray lines are improved DLCQ ($K=10$ and 100); 
and the heavy line is the exact result. \label{fig4}}
\end{figure}

So far, we have discussed the two particle bound state 
equation.  The generalization to a DLCQ calculation with more 
particles is straightforward.
For example, consider the following terms from the Schwinger 
model Hamiltonian~\cite{eller,harada}:
\begin{eqnarray}
 \mu^2 \sum_{k=1/2}^\infty \frac{ b_k^{\dag} b_k + d_k^{\dag} d_k}{k} 
\, + \, g^2 \sum_{k=1/2}^\infty \left( b_k^{\dag} b_k + d_k^{\dag} d_k\right)
   \sum_{n=1}^{k-1/2} \frac{1}{n^2} \nonumber \\
\, - \,  g^2 \! \sum_{\begin{array}{@{}c@{}}\scriptstyle k,l,m,n=1/2\\
        \scriptstyle k\ne l \end{array}}^{\infty} \!
   \frac{\delta_{k+m,l+n}}{\left(k-l \right)^2} 
   \, b_k^{\dag} b_l d_m^{\dag} d_n \label{relevant} \; .
\end{eqnarray}
In the two particle sector, this Hamiltonian produces 
the DLCQ version of the 't Hooft equation (\ref{dlcq}).
The improved DLCQ version of (\ref{relevant}) is
\begin{equation}
 \mu^2 \sum_{k=1/2}^\infty \frac{ b_k^{\dag} b_k + d_k^{\dag} d_k}{k}
\, + \, g^2 \sum_{k,l=1/2}^\infty  J_{k,l}\, b_k^{\dag} b_k d_l^{\dag} d_l
\, - \, g^2 \! \sum_{\begin{array}{@{}c@{}}\scriptstyle k,l,m,n=1/2\\
    \scriptstyle k\ne l \end{array}}^{\infty} \!
     \frac{\delta_{k+m,l+n}}{\left(k-l \right)^2}
        \, b_k^{\dag} b_l d_m^{\dag} d_n \label{imp2}
 \end{equation}
where
\begin{equation}
       J_{k,l} = J_{l,k} = {\rm I}\!\left(\frac{k}{k+l}\right)
        +  \sum_{\begin{array}{@{}c@{}}\scriptstyle j=1/2\\
	      \scriptstyle j\neq k \end{array}}^{k+l-1/2}
	   \frac{\left(\frac{j (k+l-j)}{kl}\right)^\beta}{
	   (k-j)^2} \; .
\end{equation}
We have  replaced the second term of (\ref{relevant}) by a four-point
interaction since the improved DLCQ subtraction is 
sensitive to the momenta of both particles.  The improved
form (\ref{imp2}) will remove the slow convergence
from a many particle calculation.

\section{Conclusion}

There have been some past attempts to quantify the
errors associated with DLCQ.  The claim in Ref.~\cite{hornbostel}
that the leading error is $1/K$ is certainly incorrect.  
In Ref.~\cite{tube}, periodic boundary conditions were applied to
a somewhat different theory.  In that case, a leading error of $1/\ln(K)$
seemed to explain the data.

There are a large number of DLCQ studies which have suffered
from this slow convergence problem.  This includes DLCQ studies
of (1+1)-dimensional QCD \cite{hornbostel,heyssler}, 
studies of the Schwinger model~\cite{eller,elser} (as
noted in Ref.~\cite{vary}), 
dimensionally reduced theories~\cite{tube,adjoint},
and the transverse lattice~\cite{dalley}.
However, one should note that there are many cases where
the slow convergence problem is not important:  It
is only of concern when $\mu$ is small but nonzero (the DLCQ
subtraction is correct when $\mu=0$) and $1/K$ errors can dominate
in a many particle calculation (where the size of $K$ is 
somewhat limited).

\section*{Acknowledgements}I would like to thank H.--C. Pauli
and R. Bayer for useful discussions.

\end{document}